\documentclass[preprint,showpacs,preprintnumbers,amsmath,amssymb,tightenlines,nofootinbib]{revtex4-1}
 \usepackage{graphicx}
 \usepackage{dcolumn}
 \usepackage{bm}
 \usepackage{ulem}
 \usepackage{enumitem}
 \usepackage{slashed}
 \usepackage{amssymb}
\def\q{\bm{q}}
\def\p{\bm{p}}
\def\k{\bm{k}}
\def\v{\bm{v}}

\def\l{\bm{l}}

\def\vec\epsilon{\bm{\epsilon}}

\def\q{\bm{q}}
\def\p{\bm{p}}
\def\k{\bm{k}}
\def\v{\bm{v}}

\usepackage{xcolor}

\begin{document}

\title{Nonequilibrium approach to heavy-quark transport}
\author{Juhee Hong}
\affiliation{Department of Physics and Institute of Physics and Applied Physics, Yonsei University,
Seoul 03722, Korea}
\date{\today}

\begin{abstract}

A nonequilibrium Green's function approach to heavy-quark transport is 
discussed within the framework of the Kadanoff-Baym equation. 
One- and two-loop self-energy diagrams with hard-thermal-loop resummed 
propagators are calculated in the real-time formalism. 
Under the quasiparticle approximation, the kinetic equation reduces to the  
Boltzmann equation that accounts for elastic scattering and medium-induced 
gluon emission from a single scattering. 
Numerical studies suggest that off-shell and memory effects in 
quark-gluon plasmas influence heavy-quark dynamics, particularly near 
the critical temperature. 
\end{abstract}

\maketitle

\section{Introduction}

Relativistic heavy-ion collisions create extreme matter of high temperature and 
density, which is far from equilibrium and rapidly evolves 
(for recent review, see Refs. \cite{Busza:2018rrf,Berges:2020fwq}). 
In these collision systems, heavy quarks have been extensively studied to 
understand energy loss, thermalization, and transport properties of 
quark-gluon plasmas (QGP). 
While various transport models have been developed to describe  
heavy-flavor observables which may carry signatures of nonequilibrium 
dynamics, most adopt a quasiparticle description based on a 
semiclassical Boltzmann equation. 
The Boltzmann-type transport, which relies on on-shell quasiparticles 
and Markovian dynamics, is valid for weakly coupled plasmas in the 
high-temperature and low-density limits.  
On the other hand, the strongly interacting QGP consist of unstable quarks 
and gluons, which acquire finite lifetimes and thermal masses, leading to 
broadened spectral densities.  
Moreover, quantum correlations arise from strong interactions between partons 
and their collective motion in the strongly correlated QCD matter. 
Consequently, nonequilibrium quantum dynamics is essential to describe the  
interactions between heavy quarks and the thermal medium.

Transport coefficients and experimental observables of heavy flavor have been 
evaluated by including off-shell and memory effects within phenomenological 
models 
\cite{Berrehrah:2013mua,Berrehrah:2014kba,Liu:2018syc,Liu:2020dlt,Sambataro:2020pge,Torres-Rincon:2021yga,Ruggieri:2022kxv,Pooja:2023gqt,Grishmanovskii:2025mnc}. 
While heavy-quark transport is generally described by the combined 
contribution of the collisional and radiative energy loss of heavy quarks, 
the quantum effects on inelastic scattering have not yet been fully explored. 
Because the collisional and radiative energy loss exhibit distinct momentum 
dependence in heavy-quark momentum spectra \cite{Hong:2023cwl}, 
a consistent quantum transport approach in which two energy-loss mechanisms 
are treated separately is still needed.  
In this respect, the Kadanoff-Baym equation provides a unified framework 
for extending the semiclassical Boltzmann equation and incorporating 
nonequilibrium effects in QGP. 
This paper aims to develop a quantum extension of the heavy-quark kinetic 
theory that has been established in Refs. \cite{Hong:2023cwl,Hong:2025dfj}. 
It also clarifies the approximations under which the Kadanoff-Baym equation 
reduces to the Boltzmann equation, while examining the regime of validity of 
a quasiparticle description and the impact of quantum effects on 
heavy-flavor observables.

The Kadanoff-Baym equation provides a quantum transport equation for a 
nonequilibrium Green's function. 
A brief introduction to the framework for heavy-quark transport is presented 
in Sec. \ref{kbeq}. 
In the kinetic equation, all transport processes in QGP enter 
through the heavy-quark self-energy which is constructed from interacting  
Green's functions.  
Sec. \ref{selfe} calculates one- and two-loop self-energy diagrams in 
hard-thermal-loop resummed theory. 
Within the quasiparticle approximation, the Kadanoff-Baym equation recovers the 
Boltzmann equation with elastic scattering at leading order. 
Including radiative processes, which correspond to medium-induced gluon 
emission from a single scattering, is appropriate for studying 
the transition from a kinematic regime dominated by the collisional energy 
loss to one dominated by the radiative energy loss. 
In Sec. \ref{quantum}, a generalized Kadanoff-Baym ansatz is applied to the 
kinetic equation to investigate two characteristic features of nonequilibrium 
quantum dynamics: off-shell and memory effects. 
The charm-quark interaction rates in off-shell QGP and on-shell massless QGP 
are compared, and 
the relaxation behavior of a single-particle excitation is analyzed in the 
presence of memory effects. 
Employing a running coupling constant, the temperature-dependence of the 
quantum effects is examined. 
Finally, Sec. \ref{summary} summarizes this work. 
The details on the self-energy are provided in Appendix.

\section{Kadanoff-Baym equation}
\label{kbeq}

For nonequilibrium Green's function methods, there are many references 
available  
\cite{kbbook,Danielewicz:1982kk,Chou:1984es,Landsman:1986uw,Mrowczynski:1992hq,Greiner:1998vd,Blaizot:2001nr,Cassing:2008nn}. 
The Kadanoff-Baym equation and its reduction to the semiclassical Boltzmann 
equation have been extensively discussed, 
especially in scalar field theory. 
For massive fermions, quantum transport has received growing attention 
in recent years, driven by the interest in spin polarization. 
The mass-shell constraint and the kinetic equation have been derived from the 
Schwinger-Dyson equation (for example, see Ref. \cite{Sheng:2021kfc}). 
Neglecting Poisson brackets, they can be applied to describe 
heavy-quark transport. 
This section provides a brief introduction.

In the real-time formalism, four types of two-point Green's functions are 
defined according to the time ordering on the Schwinger-Keldysh contour 
\cite{Schwinger:1960qe,Keldysh:1964ud},   
\begin{eqnarray}
S_F(x_1,x_2)&=&\langle T\psi(x_1)\bar{\psi}(x_2)\rangle \, ,
\nonumber\\
S_{\bar{F}}(x_1,x_2)&=&\langle \bar{T}\psi(x_1)\bar{\psi}(x_2)\rangle \, ,
\nonumber\\
S^>(x_1,x_2)&=&\langle \psi(x_1)\bar{\psi}(x_2)\rangle \, ,
\nonumber\\
S^<(x_1,x_2)&=&-\langle \bar{\psi}(x_2)\psi(x_1)\rangle \, ,
\end{eqnarray}
where $T$ and $\bar{T}$ are the time-ordering and antitime-ordering operators,  
respectively. 
These functions are interdependent: $S_F+S_{\bar{F}}=S^>+S^<$. 
The retarded and advanced Green's functions are related to the Wightman 
correlation functions as follows: 
\begin{eqnarray}
S_R(x_1,x_2)&=&\theta(t_1-t_2)\left[S^>(x_1,x_2)-S^<(x_1,x_2)\right] \, ,
\nonumber\\
S_A(x_1,x_2)&=&-\theta(t_2-t_1)\left[S^>(x_1,x_2)-S^<(x_1,x_2)\right] \, ,
\end{eqnarray}
which directly carry the spectral information.

In Wigner space, $S^<$ satisfies a kinetic equation that describes the 
nonequilibrium distribution function. 
Taking the Fourier transform with respect to $r=x_1-x_2$, 
the Wigner transformed function is  
\begin{equation}
S^<(X,P)=-\int d^4r \, e^{iP\cdot r} \,
\Big\langle \bar{\psi}\Big(X-\frac{r}{2}\Big)
\psi\Big(X+\frac{r}{2}\Big)\Big\rangle \, ,
\end{equation}
where $X=\frac{x_1+x_2}{2}$, and satisfies the following equation: 
\begin{equation}
P\cdot\frac{\partial}{\partial X}{\rm Tr}\left[S^<(X,P)\right]
=\frac{1}{2}{\rm Tr}\left[(\slashed{P}+m)
\left\{\Sigma^<(X,P)S^>(X,P)-\Sigma^>(X,P)S^<(X,P)\right\}\right] \, .
\end{equation}
For on-shell heavy quarks,   
the Kadanoff-Baym equation becomes\footnote{Under the quasiparticle 
approximation, $S^<(X,P)=f(X,\p)(\slashed{P}+m)\rho(P)$, where the spectral 
density of heavy quark is $\rho(P)=2\pi \, \delta(P^2-m^2)$. 
Since $\Sigma^{<,>}$ has no spin-dependent Dirac structure in isotropic and 
unpolarized plasmas, the Kadanoff-Baym equation simplifies to a scalar kinetic 
equation for the spin-averaged distribution function.}
\begin{eqnarray}
\label{kb_boltz}
\frac{P}{E_{\p}}\cdot \frac{\partial}{\partial X} f(X,\p)
&=&\frac{1}{2}\int_0^{\infty}\frac{dp^0}{2\pi} {\rm Tr}
\big[\Sigma^<(X,P) S^>(X,P)-\Sigma^>(X,P)S^<(X,P)\big] \, ,
\end{eqnarray}
where $f(X,\p)$ is the phase-space distribution function of heavy quarks.

Within the quasiparticle approximation, the Kadanoff-Baym equation reproduces 
the Boltzmann equation with elastic and inelastic scatterings. 
The leading contribution from elastic scattering arises at  
$\mathcal{O}(g^2)$ in the coupling constant.  
Inelastic scattering is suppressed by $\mathcal{O}(g^2)$ relative to elastic 
scattering in the low-momentum regime, while its contribution becomes enhanced 
by soft-momentum transfer and collinear gluon-emission as the heavy-quark 
momentum increases. 
To investigate the transition between the collisional and radiative energy loss 
of heavy quarks in the intermediate-momentum regime, both elastic scattering 
and gluon emission from a single scattering have been included in the 
Boltzmann equation \cite{Hong:2023cwl,Hong:2025dfj}. 
This work generalizes the semiclassical transport to nonequilibrium 
quantum transport.

\section{Self-energy}
\label{selfe}

In the kinetic Eq. (\ref{kb_boltz}), heavy-quark interactions in 
thermal media are incorporated into the self-energy terms on the right-hand 
side. 
Figs. \ref{selfe_lo} and \ref{selfe_nlo} show the leading- and 
next-to-leading-order self-energy diagrams which respectively result in the 
collisional and radiative energy loss of heavy quarks in the on-shell limit. 
Although heavy-quark transport receives contributions, in principle, from 
processes beyond two-body elastic scattering and medium-induced gluon 
bremsstrahlung, this paper focuses on real processes, which are relevant to 
the heavy-quark energy loss.

\subsection{Leading order}

\begin{figure}
\includegraphics[width=0.42\textwidth]{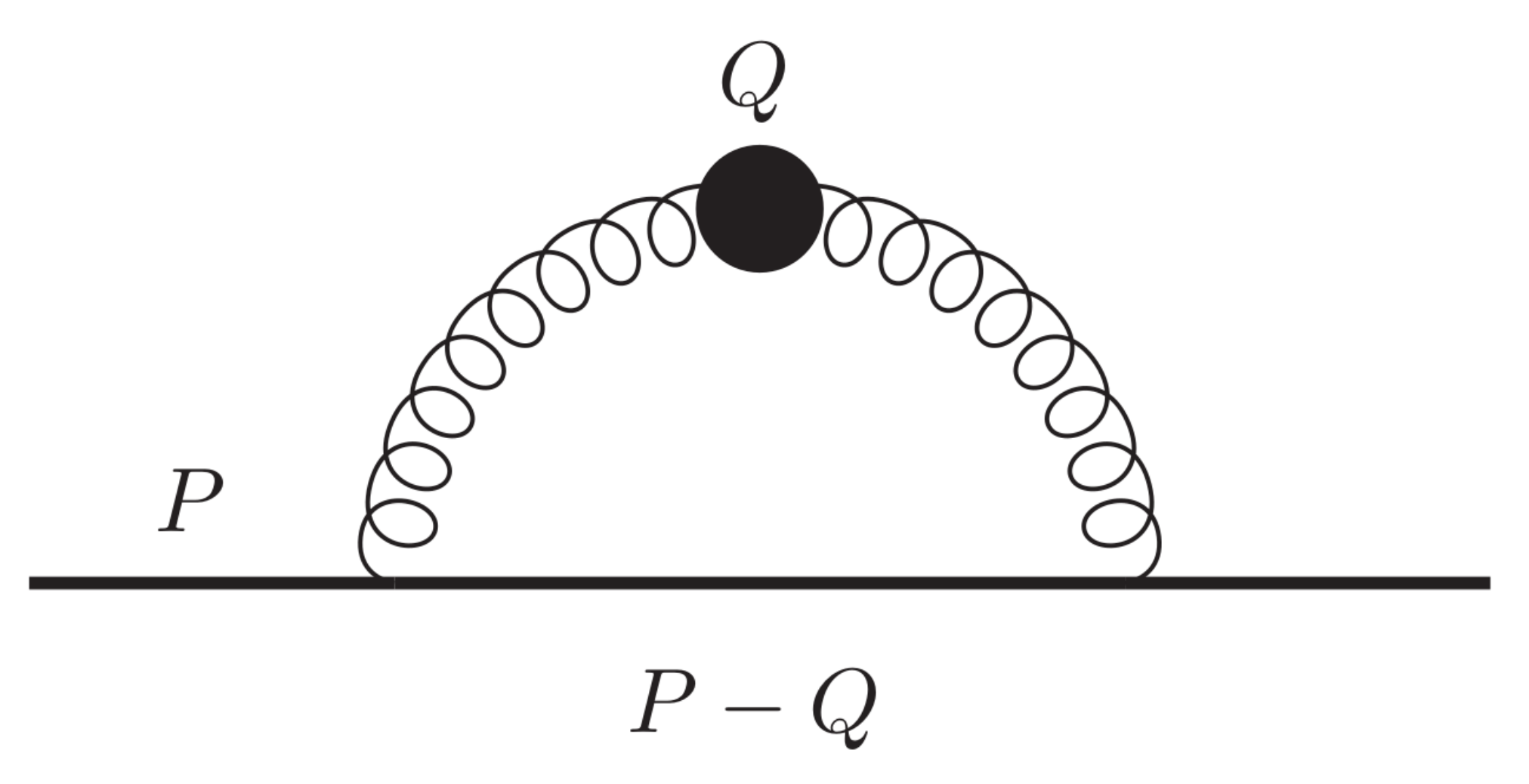}
\caption{
The heavy-quark self-energy at leading order. 
The black dot denotes the HTL resummed propagator for 
soft gluon ($Q\sim gT$). 
}
\label{selfe_lo}
\end{figure}

At leading order, the heavy-quark self-energy is given by an effective 
one-loop diagram with a hard-thermal-loop (HTL) resummed propagator, shown in 
Fig. \ref{selfe_lo} \cite{Braaten:1991jj,Braaten:1991we,Bellac:2011kqa}, 
\begin{equation}
\Sigma_{\rm LO}^>(P)
=g^2C_F\int \frac{d^4Q}{(2\pi)^4}\gamma^\mu
S^>(P-Q)\gamma^\nu G_{\mu\nu}^>(Q) \, ,
\end{equation}
where $S^>$ and $G^>_{\mu\nu}$ are the heavy-quark and gluon cut propagators, 
respectively.  
Using the Kadanoff-Baym ansatz, the heavy-quark propagators are given by 
\begin{eqnarray}
\label{hqprop}
S^>(P-Q)&=&[1-f(\p-\q)](\slashed{P}-\slashed{Q}+m)\rho(P-Q) \, ,
\nonumber\\
S^<(P-Q)&=&f(\p-\q)(\slashed{P}-\slashed{Q}+m)\rho(P-Q) \, ,
\end{eqnarray}
where the spectral density of on-shell heavy quark is 
\begin{equation}
\rho(P-Q)=\frac{1}{2E_{\p}} \, 
2\pi \, \delta(\omega-\q\cdot\v) 
\, ,
\end{equation}
with $E_{\p}=\sqrt{\p^2+m^2}$. 
For gluons with soft spacelike momentum ($Q\sim gT$), the Coulomb gauge is 
employed: 
\begin{eqnarray}
\label{gprop}
G_{\mu\nu}^>(Q)&=&[1+f(\omega)] 
\big[\rho_L(Q) g_{\mu 0}g_{\nu 0}
+\rho_T(Q)(\delta_{ij}-\hat{q}_i\hat{q}_j)\big] \, ,
\nonumber\\
G_{\mu\nu}^<(Q)&=&f(\omega)
\big[\rho_L(Q) g_{\mu 0}g_{\nu 0}
+\rho_T(Q)(\delta_{ij}-\hat{q}_i\hat{q}_j)\big] \, ,
\end{eqnarray}
where $\rho_{L,T}$ are obtained from the HTL resummed propagators 
\cite{Braaten:1989mz}, 
\begin{eqnarray}
G_{00}(Q)&=&\frac{i}{q^2+\Pi_{L}(Q)} \, ,
\nonumber\\
G_{ij}(Q)&=&\frac{i(\delta_{ij}-\hat{q}_i\hat{q}_j)}{\omega^2-q^2-\Pi_T(Q)} \, ,
\end{eqnarray}
with  
\begin{eqnarray}
\Pi_{L}(Q)&=&m_D^2\left[1-\frac{\omega}{2q}\left(\ln
\frac{q+\omega}{q-\omega}-i\pi\right)\right] \, ,
\nonumber\\
\Pi_T(Q)&=&\frac{m_D^2}{2}\left[\frac{\omega^2}{q^2}+\frac{\omega(q^2-\omega^2)}
{2q^3}\left(\ln\frac{q+\omega}{q-\omega}-i\pi\right)\right] \, .
\end{eqnarray}
Using Eq. (\ref{hqprop}), the self-energy is  
\begin{equation}
\Sigma_{\rm LO}^>(P)
=\frac{\pi g^2C_F}{E_{\p}}\int \frac{d^4Q}{(2\pi)^4} 
\delta(\omega-\q\cdot\v)\gamma^\mu
(\slashed{P}-\slashed{Q}+m)\gamma^\nu G_{\mu\nu}^>(Q)[1-f(\p-\q)] \, .
\end{equation}
Similarly, $\Sigma^<$ is determined from $\Sigma^>$ by replacing $S^>$ and 
$G^{>}$ with $S^<$ and $G^{<}$, respectively.

In the quasiparticle approximation, the right-hand side of 
Eq. (\ref{kb_boltz}) yields the collision terms for elastic scattering 
between heavy quarks and thermal particles. 
Because the heavy-quark mass is much larger than the momenta of thermal 
particles ($m\gg T$), the dominant contribution comes from 
the $t$-channel soft-gluon exchange. 
The leading-order interaction rate is 
\cite{Weldon:1983jn,Braaten:1991jj,Braaten:1991we} 
\begin{eqnarray}
\label{int_rate_lo}
\Gamma_{\rm LO}
&=&\frac{1}{4E_{\p}}{\rm Tr}\left[(\slashed{P}+m)\Sigma_{\rm LO}^>(P)\right] \, ,
\nonumber\\
&=&\frac{2\pi g^2C_F}{E_{\p}^2}
\int\frac{d^4Q}{(2\pi)^4}\delta(\omega-\q\cdot\v)[1+f(\omega)]
\left[E_{\p}^2 \, \rho_L(Q)+\{\p^2-
(\p\cdot\hat{\q})^2\}\rho_T(Q)\right] \, , \, \, \, 
\end{eqnarray}
and one of the collision terms is  
\begin{eqnarray}
\label{loss}
\frac{1}{2}\int_0^\infty\frac{dp^0}{2\pi}
{\rm Tr}\left[\Sigma_{\rm LO}^>(P)S^<(P)\right] 
&=& \Gamma_{\rm LO} f(\p)[1-f(\p-\q)] \, .
\end{eqnarray}
Note that $\delta(\omega-\q\cdot\v)$ in Eq. (\ref{int_rate_lo}) can be 
rewritten as the total energy conservation for the $P+K\rightarrow P'+K'$ 
process, where $P$ and $P'=P-Q$ are respectively heavy-quark momenta before 
and after collision, and $K$ and $K'=K+Q$ are those for thermal particle. 
The Debye screening mass in $\rho_{L,T}$ can be expressed as 
\cite{Blaizot:2001nr}
\begin{eqnarray}
\label{debye}
m_D^2&=&
-\frac{N_{c}g^2}{\pi^2}\int_0^\infty dk \, k^2\frac{df_B(k)}{dk}
-\frac{N_{f}g^2}{\pi^2}\int_0^\infty dk \, k^2\frac{df_F(k)}{dk} \, . 
\end{eqnarray}
Parameterizing 
$\int d^3\k=\int dk \, k^2\int d\cos\theta_k \int d\phi$ (where 
$\cos\theta_k=\frac{\omega}{q}$), 
changing the integration variables $\int d^3\q\rightarrow \int d^3\k'$, and 
introducing $\int d^3\p' \, \delta^3(\p+\k-\p'-\k')=1$ 
\cite{Arnold:2000dr,Moore:2004tg}, 
Eq. (\ref{loss}) becomes the loss term in the standard Boltzmann equation, 
\begin{multline}
\frac{1}{2}\int_0^\infty\frac{dp^0}{2\pi}
{\rm Tr}\left[\Sigma_{\rm LO}^>(P)S^<(P)\right] 
=\frac{1}{2E_{\p}}\int\frac{d^3\k}{(2\pi)^32k}
\int\frac{d^3\k'}{(2\pi)^32k'}\int\frac{d^3\p'}{(2\pi)^32E_{\p'}}
|\mathcal{M}|^2
\\
\times
(2\pi)^4\delta^4(P+K-P'-K')f(\p)f(k)[1-f(\p')][1\pm f(k')] \, ,
\end{multline}
where $|\mathcal{M}|^2$ is the squared matrix element, summed over color and 
spin of thermal particles, for the $t$-channel gluon exchange (see Appendix). 
Using the same procedure, the gain term in the Boltzmann equation is obtained 
from the other collision term of Eq. (\ref{kb_boltz}).

\subsection{Radiative Processes}

\begin{figure}
\includegraphics[width=0.325\textwidth]{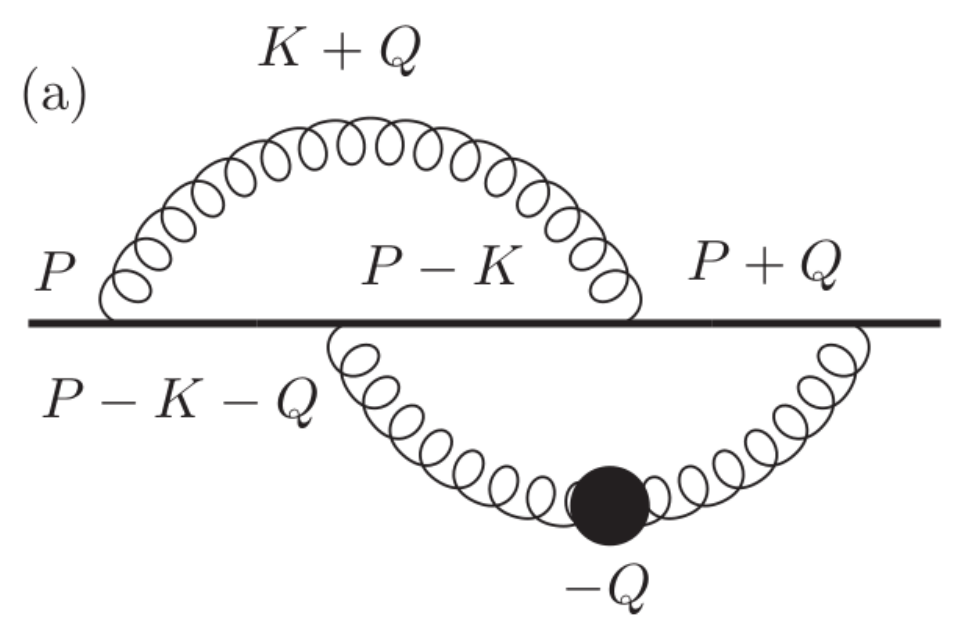}
\includegraphics[width=0.325\textwidth]{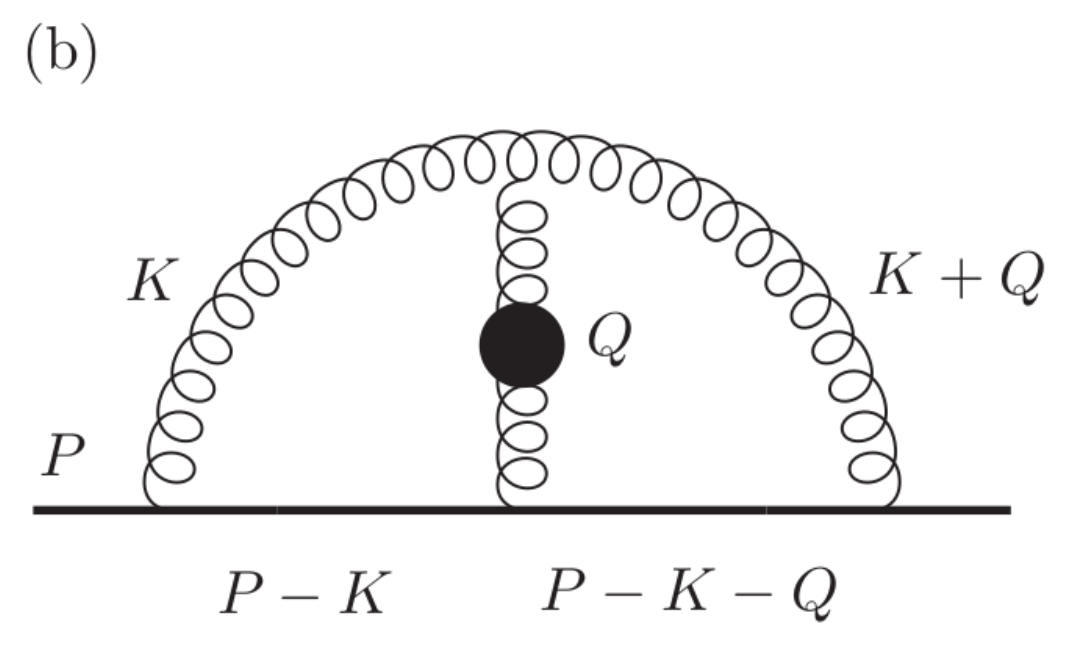}
\includegraphics[width=0.325\textwidth]{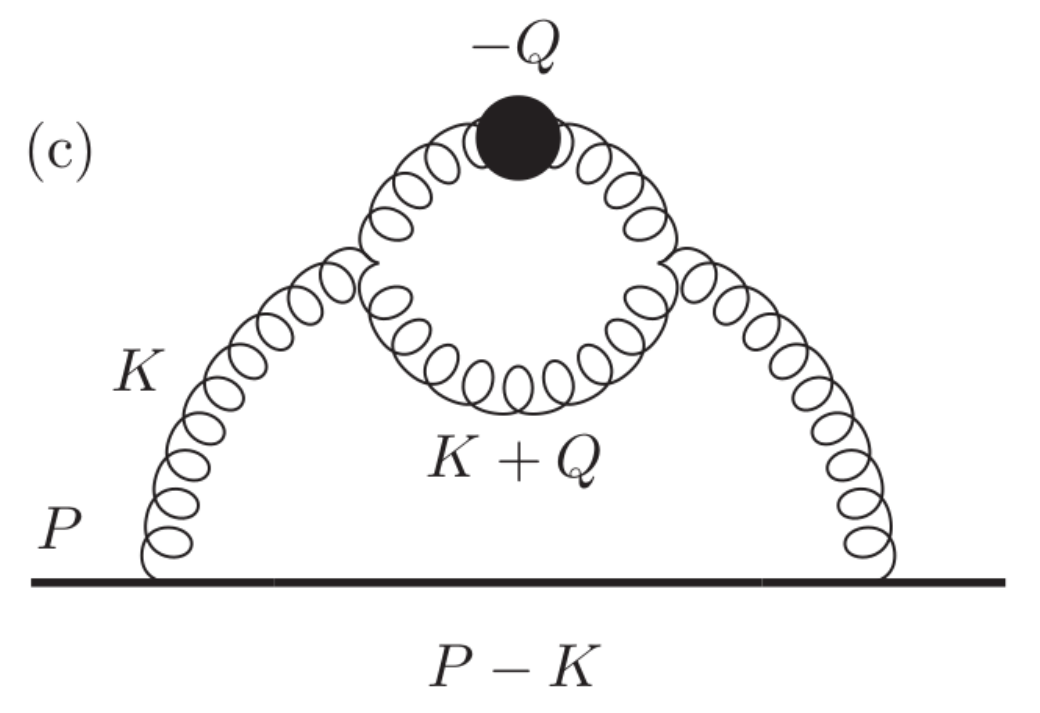}
\caption{
The self-energy diagrams for medium-induced gluon emission from a single 
scattering. 
Gluons with $K+Q$ and $K$ carry hard momenta ($K\sim T$). 
}
\label{selfe_nlo}
\end{figure}

\begin{figure}
\includegraphics[width=0.87\textwidth]{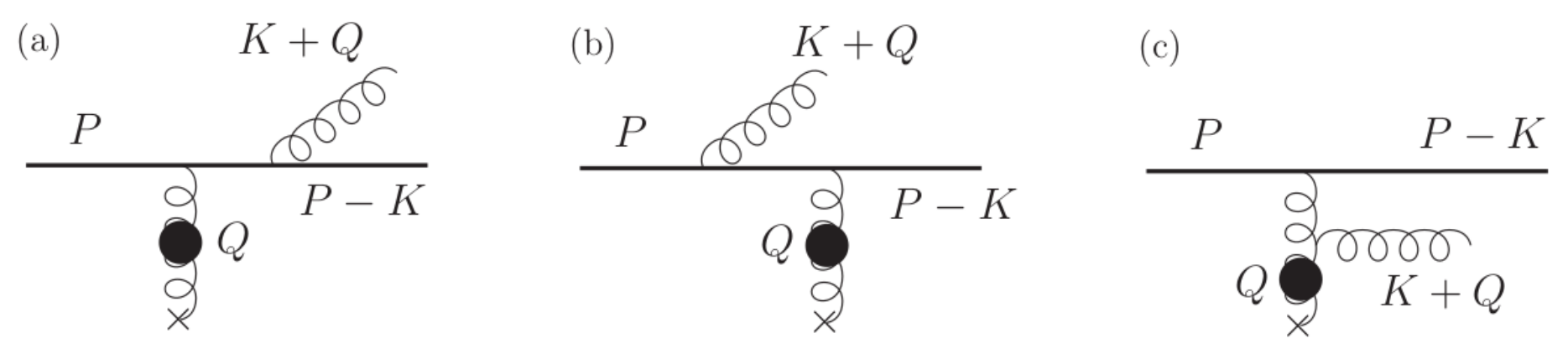}
\caption{
Medium-induced gluon emission from a single scattering. 
The crosses denote thermal scattering centers. 
}
\label{rad}
\end{figure}

Beyond leading order, numerous multi-loop self-energy diagrams contribute to 
transport processes in high-temperature QCD plasmas. 
Due to soft and collinear enhancement, multiple soft scatterings must be 
summed coherently in the ultrarelativistic limit 
\cite{Landau:1953um,Migdal:1956tc,Arnold:2001ba,Arnold:2002ja,Arnold:2002zm}. 
To investigate the transition between the collisional and radiative energy 
loss at intermediate momentum, a single-scattering regime can be 
considered for heavy-quark dynamics \cite{Hong:2023cwl}. 
Medium-induced gluon emission from a single scattering is 
described by two-loop self-energy diagrams with HTL resummed propagators, 
shown in Fig. \ref{selfe_nlo}\footnote{There are 
three additional diagrams that are similar to Fig. \ref{selfe_nlo} (a) 
(two of them with gluons uncrossed), and three additional diagrams similar 
to Fig. \ref{selfe_nlo} (b), for a total of nine diagrams corresponding to 
the squared emission processes in Fig. \ref{rad}. 
Although these diagrams have been computed previously 
\cite{Djordjevic:2007at,Djordjevic:2009cr}, this work revisits them in the 
semicollinear approximation using the heavy-quark collision kernel:  
$(2\pi)^3 \frac{d\Gamma_{\rm LO}}{d^3\q}=g^2C_F\int d\omega \, \delta(\omega-\q\cdot\v)G_>^{--}(Q)$.}. 
Since the radiative processes are effective at high momentum 
($p^z\gg m$ for heavy quark moving in $\hat{z}$), it is convenient to use the 
lightcone coordinates: 
$[p^+,p^-,\p_T]=\big[\frac{p^0+p^z}{2},p^0-p^z,\p_T\big]$.

The self-energy diagrams in Fig. \ref{selfe_nlo} are calculated in the 
semicollinear approximation\footnote{Within the semicollinear 
approximation, the formation time ($\sim \frac{2k^+}{k_T^2+m^2x^2}$) is 
shorter than the 
mean free path ($\sim\frac{1}{g^2T}$), validating gluon emission 
from a single soft scattering.}:   
$p^+\gg k^+\sim T\gg k_T\gg q^+,q_T\sim gT$ 
\cite{Ghiglieri:2013gia,Ghiglieri:2015ala,Ghiglieri:2022gyv}. 
Real radiative processes arise from cutting the $P-K$, $K+Q$, and $-Q$ 
propagators,  
\begin{eqnarray}
\label{sum_abc}
\Sigma^>_{(a)}(P) &=&
g^4T_aT_bT_aT_b  
\int\frac{d^4K}{(2\pi)^4}\int\frac{d^4Q}{(2\pi)^4}
G_>^{\mu\nu}(K+Q)G_>^{\alpha\beta}(-Q)
\nonumber\\
&&
\times\gamma_\mu S_F(P-K-Q)\gamma_\alpha
S^>(P-K)\gamma_\nu S_{\bar{F}}(P+Q)\gamma_\beta 
\, \,  + \,  \mbox{(3 others)} \, ,
\nonumber\\
\Sigma^>_{(b)}(P) &=&
-\frac{1}{2}g^4C_FC_A  
 \int\frac{d^4K}{(2\pi)^4}\int\frac{d^4Q}{(2\pi)^4}
G_F^{\mu\rho}(K)G_>^{\nu\sigma}(-Q)G_>^{\alpha\beta}(K+Q)
\nonumber\\
&&
\times 
[g_{\mu\nu}(K-Q)_\alpha+g_{\nu\alpha}(2Q+K)_\mu+g_{\alpha\mu}(-2K-Q)_\nu]
\nonumber\\
&&\times\gamma_\rho S^>(P-K)\gamma_\sigma S_{\bar{F}}(P-K-Q)\gamma_\beta 
\, \,  + \,  \mbox{(3 others)} \, ,
\nonumber\\
\Sigma^>_{(c)}(P) &=&
g^4C_FC_A  
\int\frac{d^4K}{(2\pi)^4}\int\frac{d^4Q}{(2\pi)^4}
G_F^{\lambda\mu}(K) 
G_>^{\alpha\beta}(K+Q)G_>^{\rho\sigma}(-Q)
G_{\bar{F}}^{\xi\nu}(K)
\nonumber\\
&&
\times
[g_{\alpha\mu}(-2K-Q)_\rho+g_{\mu\rho}(K-Q)_\alpha+g_{\rho\alpha}(2Q+K)_\mu]
\nonumber\\
&&
\times
[g_{\beta\sigma}(2Q+K)_\nu+g_{\sigma\nu}(K-Q)_\beta+g_{\nu\beta}(-2K-Q)_\sigma] 
\gamma_\lambda S^>(P-K)\gamma_\xi
\, .
\end{eqnarray}
To extract the leading contribution from gluon emission, the propagators of 
heavy quark and gluon are approximated as follows. 
After gluon emission, the heavy quark remains on-shell: 
\begin{eqnarray}
\label{hq_cut}
S^>(P-K)&=&
[1-f(\p-\k)] \, \frac{\slashed{P}-\slashed{K}+m}{2p^+} \, 
2\pi \, \delta\Big(k^-+\frac{m^2x^2}{2k^+}\Big) \, .
\end{eqnarray}
In the eikonal limit, the uncut heavy-quark propagators are  
\begin{eqnarray}
\label{hq_uncut}
S_F(P+Q)&=& \frac{i\gamma^-}{2 (q^-+ i\epsilon)}   \, ,
\nonumber\\
S_F(P-K-Q)&=& -\frac{i\gamma^-}{2 (q^-- i\epsilon)}   \, .
\end{eqnarray}
The uncut gluon propagator is   
\begin{eqnarray}
\label{g_uncut}
G_{F}^{ij}(K)&=&
\frac{ i(\delta^{ij}-\hat{k}^i\hat{k}^j)}{-\k_T^2-m^2x^2+ i\epsilon}\, ,
\end{eqnarray}
where $x=k^+/p^+$ and 
the $\delta$-function in Eq. (\ref{hq_cut}) has been used.  
Using Eqs. (\ref{hq_cut}), (\ref{hq_uncut}), and (\ref{g_uncut})  
and summing all terms in Eq. (\ref{sum_abc}) (see Appendix), 
\begin{eqnarray}
\label{leading_result}
\Sigma_{\rm rad}^>(P)&=&
2 g^4C_FC_A\int\frac{dk^+}{k^+}\int\frac{d^2\k_T}{(2\pi)^2}
\int\frac{d^4Q}{(2\pi)^4}
\delta\Big(q^--\frac{(\k_T+\q_T)^2+m^2x^2}{2k^+}\Big)
[1+f(k^+)]
\nonumber\\
&&\times
G_>^{--}(-Q)
\left[\frac{\k_T+\q_T}{(\k_T+\q_T)^2+m^2x^2}
-\frac{\k_T}{\k_T^2+m^2x^2}\right]^2  
[1-f(\p-\k)]\gamma_- \, .
\end{eqnarray}
Under the approximation $k_T\gg q_T$, it can be simplified as   
\begin{eqnarray}
\label{sigma_dlog}
\Sigma_{\rm rad}^>(P)&=&
2g^4C_FC_A\int\frac{dk^+}{k^+}\int\frac{d^2\k_T}{(2\pi)^2}
\int\frac{d^4Q}{(2\pi)^4}
\delta\Big(q^--\frac{\k_T^2+m^2x^2}{2k^+}\Big)
[1+f(k^+)]
\nonumber\\
&&\times
G_>^{--}(-Q)
\frac{\q_T^2}{(\k_T^2+m^2x^2)^2} [1-f(\p-\k)]\gamma_- \, .
\end{eqnarray}

With the self-energy determined, the right-hand side of the Kadanoff-Baym 
equation can be recast as the collision terms in the Boltzmann equation. 
The second term in Eq. (\ref{kb_boltz}) is  
\begin{eqnarray}
\label{Crad}
\frac{1}{2}\int_0^\infty\frac{dp^0}{2\pi}{\rm Tr }
\left[\Sigma_{\rm rad}^>(P)S^<(P)\right]
&=&
2 g^4C_FC_A\int\frac{dk^+}{k^+}\int\frac{d^2\k_T}{(2\pi)^2}
\int\frac{d^4Q}{(2\pi)^4} \,
\delta\Big(q^--\frac{\k_T^2+m^2x^2}{2k^+}\Big)
\nonumber\\
&&\times
[1+f(k^+)]G_>^{--}(-Q)
\frac{\q_T^2}{(\k_T^2+m^2x^2)^2}[1-f(\p-\k)]f(\p) \, .
\nonumber\\
\end{eqnarray}
The $\delta$-function corresponds to the total energy conservation. 
The amplitudes of radiation processes shown in Fig. \ref{rad} involve the 
$t$-channel gluon exchange with an additional heavy-quark or gluon propagator 
\cite{Gunion:1981qs},  
\begin{equation}
|\mathcal{M}_{(a)}+\mathcal{M}_{(b)}+\mathcal{M}_{(c)}|^2
=\frac{4g^2C_Aq_T^2}{(\k_T^2+m^2x^2)^2}|\mathcal{M}|^2 \, .
\end{equation}
Modifying the phase-space integration for soft scattering as described in the 
leading-order process, Eq. (\ref{Crad}) becomes  
\begin{eqnarray}
\label{Crad2}
\frac{1}{2}\int_0^\infty\frac{dp^0}{2\pi}{\rm Tr }
\left[\Sigma_{\rm rad}^>(P)S^<(P)\right]
&=&\frac{1}{2E_{\p}}
\int\frac{d^3\k}{(2\pi)^32k}
\int\frac{d^3\l}{(2\pi)^32l}
\int\frac{d^3\l'}{(2\pi)^32l'}
\int\frac{d^3\p'}{(2\pi)^32E_{\p'}}
\nonumber\\
&&\times 
|\mathcal{M}_a+\mathcal{M}_b+\mathcal{M}_c|^2
(2\pi)^4\delta^4(P+L-P'-L'-K)
\nonumber\\
&&\times
[1+f(k)] \, f(l)[1\pm f(l')][1-f(\p-\k)]f(\p) \, ,
\end{eqnarray}
which is the loss term due to radiation. 
Similarly, $\Sigma^<$ can be obtained from $\Sigma^>$ by replacing the 
propagators ($>\,\leftrightarrow \,<$ and $F\leftrightarrow\bar{F}$), 
and the corresponding self-energy term can be written as the gain term.

The collision term in Eq. (\ref{Crad2}) can be reformulated as a splitting 
process. 
Because gluon emission is nearly collinear, small transverse momenta may be 
integrated, allowing a separation in the directions of splitting 
particles \cite{Arnold:2002zm},  
\begin{eqnarray}
\frac{1}{2}\int_0^\infty\frac{dp^0}{2\pi}{\rm Tr }
\left[\Sigma_{\rm rad}^>(P)S^<(P)\right]
&=&\frac{(2\pi)^3}{2p^2}\int dk \int dp' \, \delta(p-p'-k) \, 
\gamma(\p;p'\hat{\p},k\hat{\p})
\nonumber\\
&& \qquad \qquad \qquad \qquad
\times
[1+f(k)][1-f(\p')]f(\p) \, ,
\end{eqnarray}
where $\delta(p-p'-k)$ is the longitudinal momentum conservation, and 
$\gamma(\p;p'\hat{\p},k\hat{\p})$ is the differential rate 
$\sim\frac{d\Gamma}{dp'dkdp \, d\Omega_p}$ for the splitting process, 
integrated over transverse momenta.  
Using the gluon emission rate given by \cite{Hong:2023cwl}
\begin{eqnarray}
\frac{d\Gamma}{dk}
&=&\frac{2g^2C_A}{\pi k}[1+f(k)][1-f(\p-\k)]
\int\frac{d^2\k_T}{(2\pi)^2}
\frac{\kappa_T\big\vert_{q^-=(\k_T^2+m^2x^2)/(2k)}}
{(\k_T^2+m^2x^2)^2} \, ,
\end{eqnarray}
where 
$\kappa_T=\frac{g^2C_F}{2}\int\frac{d^4Q}{(2\pi)^3} \, \delta(q^-) \, \q_T^2 \, G_>^{--}(Q)$ 
is the transverse momentum diffusion coefficient of a heavy quark, 
the loss term can be cast into \cite{Jeon:2003gi} 
\begin{eqnarray}
\frac{1}{2}\int_0^\infty\frac{dp^0}{2\pi}{\rm Tr }
\left[\Sigma_{\rm rad}^>(P)S^<(P)\right]
&=&\int dk \, \frac{d\Gamma}{dk} f(\p) \, .
\end{eqnarray}
This is the radiation term appearing in the Boltzmann equation of 
Refs. \cite{Hong:2023cwl,Hong:2025dfj}.

\section{Quantum effects}
\label{quantum}

The quasiparticle limit of the kinetic Eq. (\ref{kb_boltz}) reproduces the 
Boltzmann equation which accounts for elastic scattering and medium-induced 
gluon emission from a single scattering. 
While the Boltzmann-type equation is widely used in heavy-quark transport 
models, a more general transport approach based on the Kadanoff-Baym equation  
is necessary to examine the impact of nonequilibrium quantum dynamics. 
Building on the self-energy terms computed in Sec. \ref{selfe}, this 
section generalizes the transport equation to include off-shell and 
memory effects in QGP.

\subsection{Off-shell QGP}

In high-temperature QCD plasmas, quarks and gluons acquire an effective 
thermal mass and a finite damping rate due to strong interactions. 
The spectral densities of hard partons are broadened rather than sharply 
peaked delta functions, which modifies the heavy-quark interaction rate and 
the gluon emission rate.

The interaction rate in QGP consisting of off-shell partons is expressed 
in terms of the self-energy, where the resummed gluon propagator depends 
on the spectral functions of the partons 
[see Eqs. (\ref{spec1}) and (\ref{spec2})]. 
Since the energies of off-shell partons are not determined by their 
momenta, the rate is calculated by introducing additional energy integrations 
with spectral functions, 
\begin{multline}
\label{ps}
\Gamma_{\rm LO}
=\frac{1}{2E_{\p}}\int\frac{d^4K}{(2\pi)^4}\rho(K)
\int\frac{d^4K'}{(2\pi)^4}\rho(K')\int\frac{d^3\p'}{(2\pi)^32E_{\p'}}
|\mathcal{M}|^2
\\
\times
(2\pi)^4\delta^4(P+K-P'-K')
f(k^0)[1\pm f(k'^0)] \, .
\end{multline}
The spectral function is assumed to have a Lorentzian form 
\cite{Blaizot:2001nr,Berrehrah:2013mua,Berrehrah:2014kba}, 
\begin{equation}
\rho(K)=\frac{4\gamma k^0}{(k_0^2-\k^2-m_{g,q}^2)^2+(2\gamma k_0)^2} \, ,
\end{equation}
which satisfies the normalization condition, 
$\int_0^{\infty} \frac{dk^0}{2\pi} \, 2k^0\rho(K)=1$. 
The thermal mass of partons is $m_{g,q}\sim m_D$ and the damping rate is 
$\gamma\sim 0.2 \, g^2T$ \cite{Braaten:1990it,Braaten:1992gd}.  
Using the temperature-dependent coupling constant [$\alpha_s(2\pi T)$], 
both $m_{g,q}$ and $\gamma$ increase with temperature, with $m_{g,q}$ greater 
than $\gamma$.

The phase-space integration in Eq. (\ref{ps}) can be parameterized as 
described in Refs. \cite{Arnold:2000dr,Moore:2004tg},  
\begin{eqnarray}
\label{ps_integ}
\Gamma_{\rm LO}&=&
\frac{1}{4(2\pi)^3 E_{\p}^2}
\int dq \, q \int_{-qv}^{qv}\frac{d\omega}{v}
\int dk  
\int_0\frac{dk^0}{2\pi}k^0\rho(K)\int_0\frac{dk'^0}{2\pi}k'^0\rho(K')
\int d\cos\theta_k
\nonumber\\
&&\times
\int d\cos\theta_p \, \delta\Big(\cos\theta_p-\frac{\omega}{qv}\Big)
\int \frac{d\phi}{2\pi} \, |\mathcal{M}|^2 \, f(k^0)[1\pm f(k'^0)]
\, \delta(\omega+k'^0-k^0) \, ,
\end{eqnarray}
where $\cos\theta_k=\hat{\k}\cdot\hat{\q}$, 
$\cos\theta_p= \hat{\p}\cdot\hat{\q}$, and $\phi$ is the angle between the 
plane containing $\k,\q$ and the plane containing $\p,\q$.   
The infrared divergence in the $t$-channel gluon exchange is regulated by the 
Debye mass. 
The same phase-space parameterization is 
used to evaluate the gluon emission rate, which can be factorized into 
the underlying elastic-scattering rate and the emission factor: 
\begin{eqnarray}
\frac{d\Gamma}{dk}&\sim&
\Gamma_{\rm LO} \, 
\frac{g^2C_A}{\pi k}[1+f(k)] \int\frac{d^2\k_T}{(2\pi)^2}
\frac{q^2(1-\cos^2\theta_p)}{(\k_T^2+m^2x^2+m_g^2)^2}
\, .
\end{eqnarray}

\begin{figure}
\includegraphics[width=0.45\textwidth]{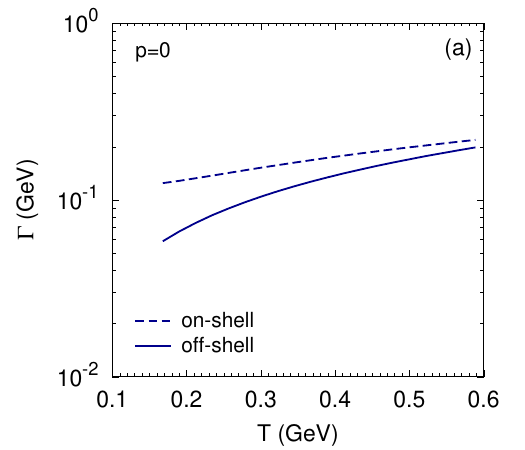}
\includegraphics[width=0.45\textwidth]{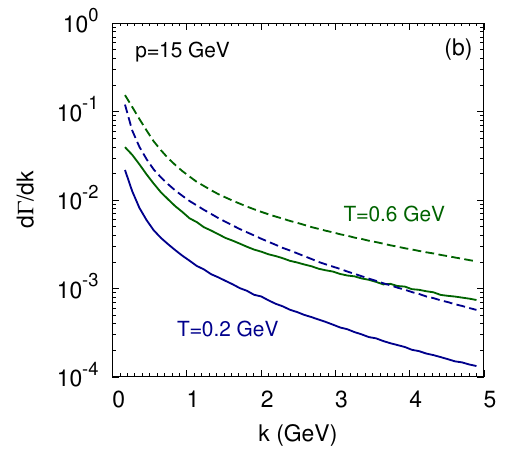}
\caption{
(a) The interaction rate of charm quark at $p=0$  
and (b) the gluon emission rate at $p=15$ GeV as a function of 
gluon momentum. 
The results for QGP of off-shell partons (solid lines) are compared with those 
for QGP of on-shell massless partons (dashed lines).  
}
\label{offshell_numeric}
\end{figure}

Figure \ref{offshell_numeric} (a) compares the interaction rate of charm quark 
at $p=0$ in QGP of off-shell partons with the rate in QGP of on-shell 
massless partons. 
The former is smaller than the latter, especially near $T_c\sim 0.157$ GeV. 
This is because the thermal mass reduces the parton densities, while 
the broadened spectral function modifies the phase space. 
Using the one-loop running coupling $\alpha_s(2\pi T)$, the coupling constant 
increases as the QGP temperature decreases, thereby enhancing the off-shell 
effects.
The gluon emission rate off charm quark at $p=15$ GeV is shown in Fig. 
\ref{offshell_numeric} (b). 
The emission rate increases with both the temperature and the charm-quark  
momentum.
Since the underlying rate for elastic scattering is reduced by the off-shell 
effects and the phase space is further suppressed by its thermal mass of 
the emitted gluon, the gluon emission rate is also lower for off-shell QGP 
than for on-shell massless QGP; this reduction is more significant at lower 
temperature.     
While the detailed effects are sensitive to the specific values 
of $m_{g,q}$, $\gamma$, and $\alpha_s$ (and typically increase with their 
magnitudes), the temperature-dependence of the off-shell 
effects is expected to remain qualitatively unchanged.

The interaction rate and transport coefficients of heavy quarks have been 
computed within dynamical quasiparticle models, which demonstrate that 
a finite mass and damping rate reduce the rate while its magnitude depends 
strongly on parameters such as the infrared regulator and the coupling 
constant \cite{Berrehrah:2013mua,Berrehrah:2014kba,Sambataro:2020pge}. 
Lower interaction rate may decrease the drag and momentum diffusion 
coefficients, thereby suppressing the energy loss of heavy quarks.   
To refine the analysis in phenomenological studies, more realistic 
parameter values and spectral functions must be specified prior to computing 
the rate.

\subsection{Memory effects}

In strongly coupled QCD plasmas, the collision time may not be much shorter 
than the mean time between collisions. 
Then heavy-quark interactions are nonlocal in time, and non-Markovian dynamics 
may be more appropriate to describe transport processes. 
When correlations persist over time and significantly affect later dynamics, 
memory effects can be important.   
This section investigates the influence of memory effects on heavy-quark 
relaxation in spatially uniform QGP.

Memory effects can be incorporated into the kinetic equation through 
nonlocal collision terms.
Integrating over past times of two-time Green's functions, 
Eq. (\ref{kb_boltz}) becomes 
\begin{eqnarray}
\label{non-markovian}
\frac{\partial}{\partial t}f(t,\p)&=&
\frac{1}{2}\int_{t_0}^t dt' \int_0^{\infty} \frac{dp^0}{2\pi} \, 
{\rm Tr}\big[\Sigma^<(t,t',P)S^>(t',t,P)
-\Sigma^>(t,t',P)S^<(t',t,P)\big] \, . 
\end{eqnarray}
Solving the Kadanoff-Baym equation for a two-time Green's function is 
computationally demanding \cite{Juchem:2003bi}. 
A generalized Kadanoff-Baym ansatz substantially simplifies the problem while 
preserving its non-Markovian nature \cite{Lipavsky:1986zz}.

To investigate how memory effects modify relaxation processes, 
a single-particle excitation is considered by adding a heavy 
quark with energy $E_{\p}$ and momentum $\p$ at $t_0=0$ to a system initially 
in equilibrium \cite{Blaizot:2001nr}. 
Using a causal generalized Kadanoff-Baym ansatz 
\cite{Danielewicz:1982kk,Greiner:1994xm}, 
\begin{eqnarray}
S^>(t,t',\p)&=&[1-f(t',\p)] \,
\frac{\slashed{P}+m}{2E_{\p}} \, e^{-iE_{\p}(t-t')}
\, ,
\nonumber\\
S^<(t,t',\p)&=&f(t',\p) \,
\frac{\slashed{P}+m}{2E_{\p}} \, e^{-iE_{\p}(t-t')}
\, ,
\end{eqnarray}
and 
\begin{eqnarray}
G_{\mu\nu}^>(t,t',\k)&=&[1+f(t',\k)] \, \frac{1}{2E_{\k}} \,
\big[g_{\mu 0}g_{\nu 0}
+(\delta_{ij}-\hat{k}_i\hat{k}_j)\big] \, 
e^{-iE_{\k}(t-t')} \, ,
\nonumber\\
G_{\mu\nu}^<(t,t',\k)&=&f(t',\k) \, \frac{1}{2E_{\k}} \,
\big[g_{\mu 0}g_{\nu 0}
+(\delta_{ij}-\hat{k}_i\hat{k}_j)\big] \, 
e^{-iE_{\k}(t-t')} \, ,
\end{eqnarray}
where the distribution functions are evaluated at the earlier time $t'$. 
To leading order in perturbation, $f$ in the self-energies is replaced by 
$f_{\rm eq}$  [$\Sigma^<\sim\Gamma f_{\rm eq}$ and 
$\Sigma^>\sim\Gamma(1-f_{\rm eq})$], linearizing Eq. (\ref{non-markovian}) as 
\cite{Blaizot:2001nr,Ikeda:2004in}
\begin{equation}
\frac{\partial}{\partial t}\delta f(t)
=-\Gamma(t) \, \delta f(0)
-\int_0^t dt' \, \Gamma(t-t')\frac{\partial}{\partial t'}\delta f(t') \, ,
\end{equation}
where $\delta f=f-f_{\rm eq}$ and 
$\Gamma(t)$ is the time-dependent interaction rate. 
For $p=0$, 
\begin{eqnarray}
\Gamma(t) 
&=&2\pi g^2C_FT\int\frac{d^3\q}{(2\pi)^3}\frac{m_D^2}{q(q^2+m_D^2)^2}
\int\frac{d\omega}{2\pi}\frac{\sin[(\omega+E_{\p-\q}-E_{\p})t]}{\omega+E_{\p-\q}-E_{\p}} \, ,
\end{eqnarray}
where the integrations over $(\omega,\q)$ are dominated by soft momentum 
$\sim gT$.  
In the limit $t\rightarrow\infty$, 
$\frac{\sin[(\omega+E_{\p-\q}-E_{\p})t]}{\omega+E_{\p-\q}-E_{\p}}\rightarrow \pi\delta(\omega-\q\cdot\v)$ 
which corresponds to the energy conservation in the local collision terms in 
the Boltzmann equation; the excitation decays exponentially with  
$\Gamma(t\rightarrow \infty)=\frac{g^2C_FT}{4\pi}$ \cite{Pisarski:1993rf}.

\begin{figure}
\includegraphics[width=0.45\textwidth]{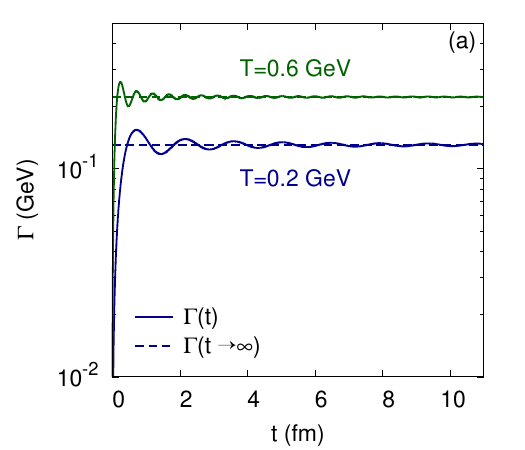}
\includegraphics[width=0.45\textwidth]{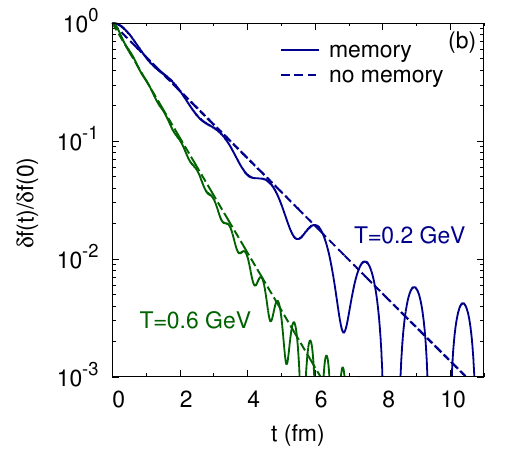}
\caption{
(a) The time-dependent interaction rate of charm quark at $p=0$.   
(b) Comparison of the relaxation of a single-particle excitation 
with (solid lines) and without (dashed lines) memory effects. 
}
\label{excit}
\end{figure}

Figure \ref{excit} (a) shows the time-dependent interaction rate of charm 
quark at $p=0$. 
The rate starts at zero and increases at early times. 
As $t$ increases, $\Gamma(t)$ exhibits damped oscillations about 
the $t\rightarrow\infty$ limit. 
Using the temperature-dependent coupling constant [$\alpha_s(2\pi T)$], 
both the interaction rate and the oscillating frequency 
increase with the QGP temperature.  
Fig. \ref{excit} (b) compares the relaxation of a single-particle excitation 
with and without memory effects. 
Since the interaction rate at $T=0.6$ GeV is higher than the rate at $T=0.2$ 
GeV, the excitation decays faster at $T=0.6$ GeV. 
When memory effects are present, the excitation relaxes 
nontrivially, developing oscillations. 
A deviation from exponential decay is observed at late times.  
The magnitude of the numerical solution with memory effects can exceed that 
without memory effects; memory effects result in slower relaxation 
\cite{Ruggieri:2022kxv,Pooja:2023gqt}. 
This reduction in the relaxation rate is more pronounced at lower temperature, 
where the coupling is stronger.

Memory effects on heavy-quark dynamics have been investigated in the 
framework of the Langevin equation driven by temporally correlated noise 
\cite{Ruggieri:2022kxv,Pooja:2023gqt}. 
They are found to reduce the heavy-quark energy loss and slow the 
evolution of momentum broadening, resulting in a longer thermalization time, 
in qualitative agreement with the present study. 
The memory effects can modify heavy-flavor observables such as the 
nuclear modification factor and the elliptic flow. 
The suppression factor is well described by heavy-quark transport models, 
whereas the elliptic flow is typically underestimated. 
It has been suggested that the memory effects may help improve the agreement 
with experimental data by enhancing the anisotropic flow while reproducing 
the suppression factor.

\section{Summary}
\label{summary}

This work discusses a nonequilibrium Green's function approach to heavy-quark 
transport. 
The Kadanoff-Baym equation with heavy-quark self-energy 
provides a useful framework for describing quantum transport while 
consistently incorporating both the collisional and radiative energy loss 
of heavy quarks in strongly correlated QCD plasmas. 
The self-energy diagrams, which include elastic scattering and medium-induced 
gluon emission from a single scattering, have been computed up to two-loop 
order in HTL resummed theory. 
It has been shown that, under the quasiparticle approximation, the kinetic 
equation reduces to the Boltzmann equation to describe heavy-quark diffusion 
and radiation.

The influence of quantum effects on heavy-quark dynamics has been 
investigated by applying a generalized Kadanoff-Baym ansatz to the self-energy 
terms. 
The numerical results indicate that the elastic scattering rate and the 
gluon emission rate in QGP of off-shell partons are lower than those in QGP of 
on-shell massless partons. 
When the mean free path becomes comparable to the correlation length in 
strongly coupled plasmas, memory effects can reduce the relaxation rate.  
These quantum effects are more pronounced at lower temperature 
where the coupling constant becomes larger, whereas the semiclassical 
Boltzmann equation is applicable in the high-temperature regime.

In strongly coupled QCD matter in the vicinity of the phase transition and 
hadronization, nonperturbative quantum effects are expected to be significant 
and consequently influence heavy-flavor observables in relativistic 
heavy-ion collisions. 
This work presents how to integrate quantum effects into heavy-quark transport 
based on the self-energy built from nonequilibrium Green's functions. 
Although the generality of the Kadanoff-Baym equation necessitates  
approximations in numerical computations, this approach enables the 
exploration of quantum kinetic theory for heavy quarks. 
Since off-shell and memory effects are closely connected and both 
depend on the time-dependent spectral functions, they should be treated in 
parallel and analyzed simultaneously. 
Furthermore, additional features of nonequilibrium quantum dynamics, 
including quantum coherence and nonlocality in space, should also be taken 
into account. 
A more comprehensive study of quantum transport will be pursued in future 
work.

\appendix

\section{Details on self-energy}

For the $t$-channel elastic scattering, the squared matrix elements are 
as follows: 
\begin{eqnarray}
|\mathcal{M}|_{q}^2&=&
16g^4C_FN_f\left[
\frac{E_{\p}^2(k^0k'^0+k^2+m_qm_q')}{|q^2+\Pi_{L}|^2}
\right.
\nonumber\\
&&
\left.+
\frac{2\{\p\cdot\k-(\p\cdot\hat{\q})(\k\cdot\hat{\q})\}^2
+(k^0k'^0-k^2-m_qm_q')\{\p^2-(\p\cdot\hat{\q})^2\}}{|q^2-\omega^2+\Pi_T|^2}
\right] \, ,
\nonumber\\
|\mathcal{M}|_{g}^2&=&
16g^4C_FN_c\left[\frac{E_{\p}^2(k^0+k'^0)^2}{2|q^2+\Pi_{L}|^2}
+
\frac{2\{\p\cdot\k-(\p\cdot\hat{\q})(\k\cdot\hat{\q})\}^2}
{|q^2-\omega^2+\Pi_T|^2}\right] \, .
\end{eqnarray}

In the high-momentum limit ($p\gg m$), the interaction rate in Eq. 
(\ref{int_rate_lo}) approaches the leading-order soft-scattering rate for a 
jet parton \cite{Arnold:2001ba}, 
\begin{eqnarray}
\Gamma_{\rm LO}
&=&g^2C_F\int\frac{d^3\q}{(2\pi)^3}
\int d\omega \, \delta(\omega-\q\cdot\v) G_{>}^{--}(Q) \, ,
\end{eqnarray}
where 
\begin{eqnarray}
G_{>}^{--}(Q)
&=&[1+f(\omega)]\Big[\rho_L(Q)+\frac{\q_T^2}{q^2}\rho_T(Q)\Big] 
\, .
\end{eqnarray}

\begin{figure}
\includegraphics[width=0.325\textwidth]{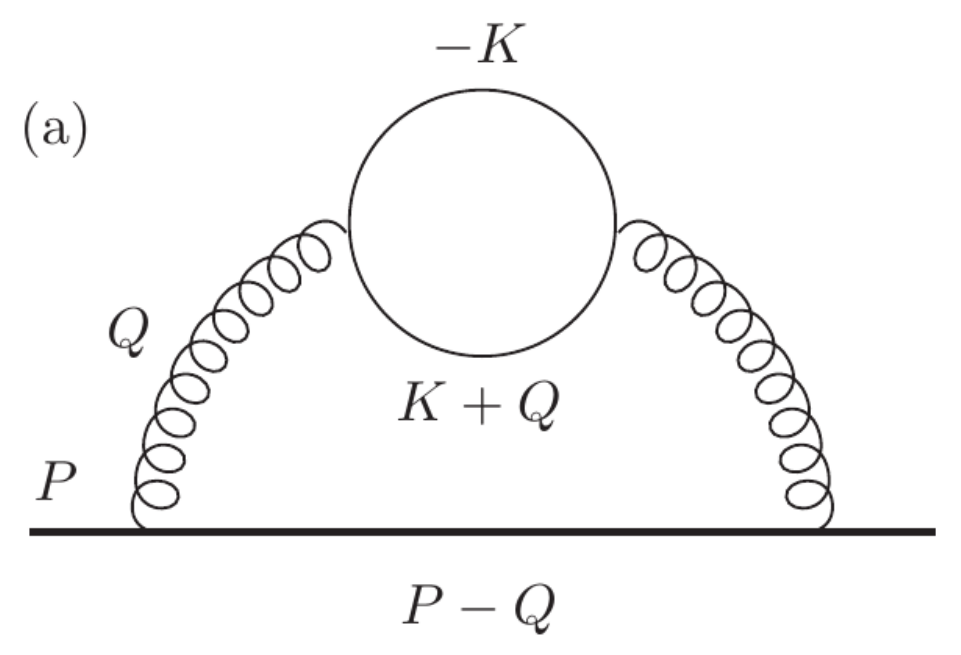}
\qquad
\qquad
\qquad
\includegraphics[width=0.325\textwidth]{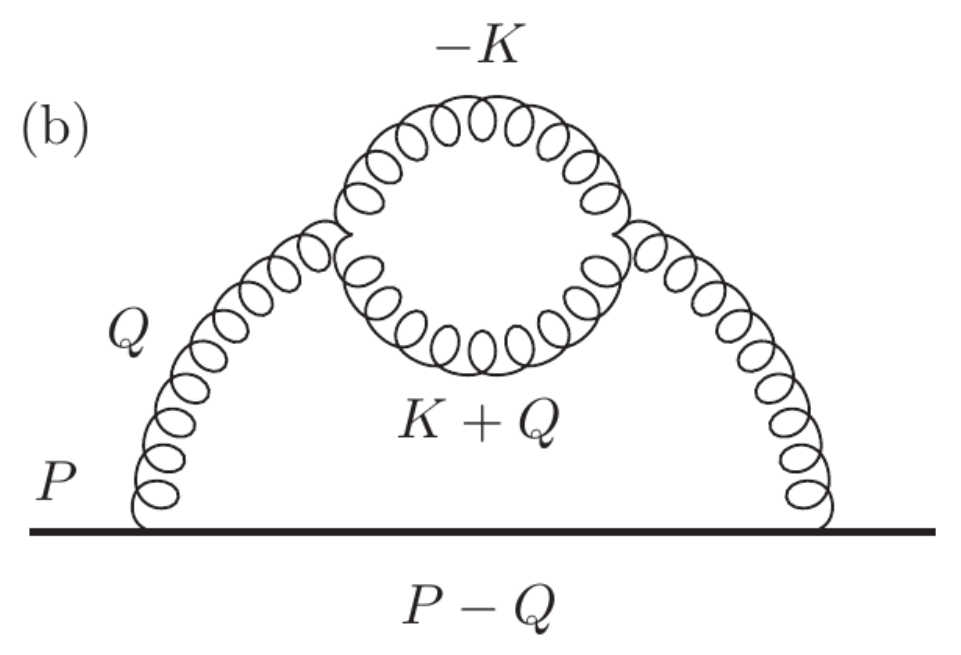}
\caption{
Quark- and gluon-loop contributions in the leading-order self-energy. 
}
\label{two_loop}
\end{figure}

The resummed gluon propagator has quark- and gluon-loop contributions 
(see Fig. \ref{two_loop}):  
\begin{eqnarray}
\label{spec1}
G_>^{\mu\nu}(Q)=
N_f g^2\int\frac{d^4K}{(2\pi)^4} 
{\rm Tr}\big[\gamma_{\alpha}S^>(-K)\gamma_{\beta}S^>(K+Q)\big] 
G_F^{\mu\alpha}(Q)G_{\bar{F}}^{\nu\beta}(Q) 
\, \, + \, (g\mbox{--loop}) \, . 
\end{eqnarray}
The quark propagators are  
\begin{eqnarray}
\label{spec2}
S^>(-K)&=&f(k)(\slashed{K}+m_q)\rho(K) \, 
\, \rightarrow \, \,
f(k)\frac{\slashed{K}}{2k}\, 2\pi \, \delta(k^0-k)\, ,
\nonumber\\
S^>(K+Q)&=&[1-f(k')](\slashed{K'}+m_q')\rho(K') \, 
\, \rightarrow \, \,
[1-f(k')]\frac{\slashed{K}}{2kq} \, 2\pi \,
\delta\Big(\cos\theta_k-\frac{\omega}{q}\Big)
 \, ,
\end{eqnarray}
where $\rightarrow$ denotes the quasiparticle limit.

Using the heavy-quark propagators [Eqs. (\ref{hq_cut}) and (\ref{hq_uncut})], 
the self-energies in Eq. (\ref{sum_abc}) become 
\begin{eqnarray}
\label{combine}
\Sigma^>_{(a)}(P) &=&
g^4C_FC_A\int\frac{d^4K}{(2\pi)^4}\int\frac{d^4Q}{(2\pi)^4}
G^{--}_>(K+Q)G^{--}_>(-Q)\frac{1}{(q^-)^2}
\nonumber\\
&&\times
2\pi \, \delta\Big(k^-+\frac{m^2x^2}{2k^+}\Big)
[1-f(\p-\k)]\gamma_- \, , 
\nonumber\\
\Sigma^>_{(b)}(P) &=&
-2ig^4C_FC_A \int\frac{d^4K}{(2\pi)^4}\int\frac{d^4Q}{(2\pi)^4}
G_F^{\mu-}(K)G_>^{\nu-}(-Q)G_>^{\alpha-}(K+Q)\frac{1}{q^-}
\nonumber\\
&&
\times
[g_{\mu\nu}(K-Q)_\alpha+g_{\nu\alpha}(2Q+K)_\mu+g_{\alpha\mu}(-2K-Q)_\nu]
\nonumber\\
&&\times
2\pi \, \delta\Big(k^-+\frac{m^2x^2}{2k^+}\Big)
[1-f(\p-\k)]\gamma_- \, ,
\nonumber\\
\Sigma^>_{(c)}(P) &=&
g^4C_FC_A\int\frac{d^4K}{(2\pi)^4}\int\frac{d^4Q}{(2\pi)^4}
G_F^{-\mu}(K) 
G_>^{\alpha\beta}(K+Q)G_>^{\rho\sigma}(-Q)
G_{\bar{F}}^{-\nu}(K)
\nonumber\\
&&
\times
[g_{\alpha\mu}(-2K-Q)_\rho+g_{\mu\rho}(K-Q)_\alpha+g_{\rho\alpha}(2Q+K)_\mu]
\nonumber\\
&&
\times
[g_{\beta\sigma}(2Q+K)_\nu+g_{\sigma\nu}(K-Q)_\beta+g_{\nu\beta}(-2K-Q)_\sigma] 
\nonumber\\
&&\times
2\pi \, \delta\Big(k^-+\frac{m^2x^2}{2k^+}\Big)
[1-f(\p-\k)]
\gamma_- \, .
\end{eqnarray}
Since the emitted gluon is transverse, the gluon propagators are approximated 
as  
\begin{eqnarray}
G_>^{--}(K+Q)&=&\frac{(\k_T+\q_T)^2}{(k^+)^2}\rho(K+Q)[1+f(k^+)] \, ,
\nonumber\\
G_>^{i-}(K+Q)&=&\frac{k^i+q^i}{k^+}\rho(K+Q)[1+f(k^+)] \, ,
\nonumber\\
G_>^{ij}(K+Q)&=&\delta^{ij}\rho(K+Q)[1+f(k^+)] \, ,
\nonumber\\
G_{F}^{i-}(K)&=&\frac{k^i}{k^+}\frac{i}{2k^+k^--\k_T^2} \, ,
\end{eqnarray}
where $i,j=x,y$ and the spectral density is given by  
\begin{eqnarray}
\rho(K+Q)= \frac{1}{2k^+}\, 2\pi \, \delta\Big(q^--\frac{(\k_T+\q_T)^2+m^2x^2}{2k^+}\Big) \, , 
\end{eqnarray}
where the $\delta$-function in Eq. (\ref{hq_cut}) has been used.  
Substituting these expressions and collecting all terms in Eq. (\ref{combine}) 
yields Eq. (\ref{leading_result}).

\section*{Acknowledgments}

I would like to thank Sangyong Jeon for useful comments. 
This work was supported by the National Research Foundation of Korea(NRF) 
grant funded by the Korea government(MSIT) (RS-2024-00342514).


\begin{thebibliography}{99}

\bibitem{Busza:2018rrf}
W.~Busza, K.~Rajagopal and W.~van der Schee,
Ann. Rev. Nucl. Part. Sci. \textbf{68}, 339-376 (2018)
[arXiv:1802.04801 [hep-ph]].

\bibitem{Berges:2020fwq}
J.~Berges, M.~P.~Heller, A.~Mazeliauskas and R.~Venugopalan,
Rev. Mod. Phys. \textbf{93}, no.3, 035003 (2021)
[arXiv:2005.12299 [hep-th]].

\bibitem{Berrehrah:2013mua}
H.~Berrehrah, E.~Bratkovskaya, W.~Cassing, P.~B.~Gossiaux, J.~Aichelin and M.~Bleicher,
Phys. Rev. C \textbf{89}, no.5, 054901 (2014)
[arXiv:1308.5148 [hep-ph]].

\bibitem{Berrehrah:2014kba}
H.~Berrehrah, P.~B.~Gossiaux, J.~Aichelin, W.~Cassing and E.~Bratkovskaya,
Phys. Rev. C \textbf{90}, no.6, 064906 (2014)
[arXiv:1405.3243 [hep-ph]].

\bibitem{Liu:2018syc}
S.~Y.~F.~Liu, M.~He and R.~Rapp,
Phys. Rev. C \textbf{99}, no.5, 055201 (2019)
[arXiv:1806.05669 [nucl-th]].

\bibitem{Liu:2020dlt}
S.~Y.~F.~Liu and R.~Rapp,
JHEP \textbf{08}, 168 (2020)
[arXiv:2003.12536 [nucl-th]].

\bibitem{Sambataro:2020pge}
M.~L.~Sambataro, S.~Plumari and V.~Greco,
Eur. Phys. J. C \textbf{80}, no.12, 1140 (2020)
[arXiv:2005.14470 [hep-ph]].

\bibitem{Torres-Rincon:2021yga}
J.~M.~Torres-Rincon, G.~Monta{\~n}a, {\`A}.~Ramos and L.~Tolos,
Phys. Rev. C \textbf{105}, no.2, 025203 (2022)
[arXiv:2106.01156 [hep-ph]].

\bibitem{Ruggieri:2022kxv}
M.~Ruggieri, Pooja, J.~Prakash and S.~K.~Das,
Phys. Rev. D \textbf{106}, no.3, 034032 (2022)
[arXiv:2203.06712 [hep-ph]].

\bibitem{Pooja:2023gqt}
Pooja, S.~K.~Das, V.~Greco and M.~Ruggieri,
Phys. Rev. D \textbf{108}, no.5, 054026 (2023)
[arXiv:2306.13749 [hep-ph]].

\bibitem{Grishmanovskii:2025mnc}
I.~Grishmanovskii, T.~Song, C.~Greiner and E.~Bratkovskaya,
Phys. Rev. D \textbf{112}, no.1, 014042 (2025)
[arXiv:2503.22311 [hep-ph]].

\bibitem{Hong:2023cwl}
J.~Hong,
Phys. Rev. C \textbf{109}, no.2, 024913 (2024)
[arXiv:2308.14530 [hep-ph]].

\bibitem{Hong:2025dfj}
J.~Hong,
Phys. Rev. C \textbf{111}, no.4, 044910 (2025)
[arXiv:2501.01600 [hep-ph]].

\bibitem{kbbook}
L. P. Kadanoff and G. Baym, 
``Quantum Statistical Mechanics,'' 
Benjamin, New York, 1962.

\bibitem{Danielewicz:1982kk}
P.~Danielewicz,
Annals Phys. \textbf{152}, 239-304 (1984).

\bibitem{Chou:1984es}
K.~c.~Chou, Z.~b.~Su, B.~l.~Hao and L.~Yu,
Phys. Rept. \textbf{118}, 1-131 (1985). 

\bibitem{Landsman:1986uw}
N.~P.~Landsman and C.~G.~van Weert,
Phys. Rept. \textbf{145}, 141 (1987).

\bibitem{Mrowczynski:1992hq}
S.~Mrowczynski and U.~W.~Heinz,
Annals Phys. \textbf{229}, 1-54 (1994). 

\bibitem{Greiner:1998vd}
C.~Greiner and S.~Leupold,
Annals Phys. \textbf{270}, 328-390 (1998)
[arXiv:hep-ph/9802312 [hep-ph]].

\bibitem{Blaizot:2001nr}
J.~P.~Blaizot and E.~Iancu,
Phys. Rept. \textbf{359}, 355-528 (2002)
[arXiv:hep-ph/0101103 [hep-ph]].

\bibitem{Cassing:2008nn}
W.~Cassing,
Eur. Phys. J. ST \textbf{168}, 3-87 (2009)
[arXiv:0808.0715 [nucl-th]].

\bibitem{Sheng:2021kfc}
X.~L.~Sheng, N.~Weickgenannt, E.~Speranza, D.~H.~Rischke and Q.~Wang,
Phys. Rev. D \textbf{104}, no.1, 016029 (2021)
[arXiv:2103.10636 [nucl-th]].

\bibitem{Schwinger:1960qe}
J.~S.~Schwinger,
J. Math. Phys. \textbf{2}, 407-432 (1961).

\bibitem{Keldysh:1964ud}
L.~V.~Keldysh,
Sov. Phys. JETP \textbf{20}, 1018-1026 (1965).

\bibitem{Braaten:1991jj}
E.~Braaten and M.~H.~Thoma,
Phys. Rev. D \textbf{44}, 1298 (1991).

\bibitem{Braaten:1991we}
E.~Braaten and M.~H.~Thoma,
Phys. Rev. D \textbf{44}, R2625 (1991).

\bibitem{Bellac:2011kqa}
M.~L.~Bellac,
``Thermal Field Theory,''
Cambridge University Press, 2011.

\bibitem{Braaten:1989mz}
E.~Braaten and R.~D.~Pisarski,
Nucl. Phys. B \textbf{337}, 569-634 (1990).

\bibitem{Weldon:1983jn}
H.~A.~Weldon,
Phys. Rev. D \textbf{28}, 2007 (1983).

\bibitem{Arnold:2000dr}
P.~B.~Arnold, G.~D.~Moore and L.~G.~Yaffe,
JHEP \textbf{11}, 001 (2000)
[arXiv:hep-ph/0010177 [hep-ph]].

\bibitem{Moore:2004tg}
G.~D.~Moore and D.~Teaney,
Phys. Rev. C \textbf{71}, 064904 (2005)
[arXiv:hep-ph/0412346 [hep-ph]].

\bibitem{Landau:1953um}
L.~D.~Landau and I.~Pomeranchuk,
Dokl. Akad. Nauk Ser. Fiz. \textbf{92}, 535-536 (1953).

\bibitem{Migdal:1956tc}
A.~B.~Migdal,
Phys. Rev. \textbf{103}, 1811-1820 (1956).

\bibitem{Arnold:2001ba}
P.~B.~Arnold, G.~D.~Moore and L.~G.~Yaffe,
JHEP \textbf{11}, 057 (2001)
[arXiv:hep-ph/0109064 [hep-ph]].

\bibitem{Arnold:2002ja}
P.~B.~Arnold, G.~D.~Moore and L.~G.~Yaffe,
JHEP \textbf{06}, 030 (2002)
[arXiv:hep-ph/0204343 [hep-ph]].

\bibitem{Arnold:2002zm}
P.~B.~Arnold, G.~D.~Moore and L.~G.~Yaffe,
JHEP \textbf{01}, 030 (2003)
[arXiv:hep-ph/0209353 [hep-ph]].

\bibitem{Djordjevic:2007at}
M.~Djordjevic and U.~Heinz,
Phys. Rev. C \textbf{77}, 024905 (2008)
[arXiv:0705.3439 [nucl-th]].

\bibitem{Djordjevic:2009cr}
M.~Djordjevic,
Phys. Rev. C \textbf{80}, 064909 (2009)
[arXiv:0903.4591 [nucl-th]].

\bibitem{Ghiglieri:2013gia}
J.~Ghiglieri, J.~Hong, A.~Kurkela, E.~Lu, G.~D.~Moore and D.~Teaney,
JHEP \textbf{05}, 010 (2013)
[arXiv:1302.5970 [hep-ph]].

\bibitem{Ghiglieri:2015ala}
J.~Ghiglieri, G.~D.~Moore and D.~Teaney,
JHEP \textbf{03}, 095 (2016)
[arXiv:1509.07773 [hep-ph]].

\bibitem{Ghiglieri:2022gyv}
J.~Ghiglieri and E.~Weitz,
JHEP \textbf{11}, 068 (2022)
[arXiv:2207.08842 [hep-ph]].

\bibitem{Gunion:1981qs}
J.~F.~Gunion and G.~Bertsch,
Phys. Rev. D \textbf{25}, 746 (1982).

\bibitem{Jeon:2003gi}
S.~Jeon and G.~D.~Moore,
Phys. Rev. C \textbf{71}, 034901 (2005)
[arXiv:hep-ph/0309332 [hep-ph]].

\bibitem{Braaten:1990it}
E.~Braaten and R.~D.~Pisarski,
Phys. Rev. D \textbf{42}, 2156-2160 (1990). 

\bibitem{Braaten:1992gd}
E.~Braaten and R.~D.~Pisarski,
Phys. Rev. D \textbf{46}, 1829-1834 (1992).

\bibitem{Juchem:2003bi}
S.~Juchem, W.~Cassing and C.~Greiner,
Phys. Rev. D \textbf{69}, 025006 (2004)
[arXiv:hep-ph/0307353 [hep-ph]].

\bibitem{Lipavsky:1986zz}
P.~Lipavsky, V.~Spicka and B.~Velicky,
Phys. Rev. B \textbf{34}, 6933-6942 (1986).

\bibitem{Greiner:1994xm}
C.~Greiner, K.~Wagner and P.~G.~Reinhard,
Phys. Rev. C \textbf{49}, 1693-1701 (1994). 

\bibitem{Ikeda:2004in}
T.~Ikeda,
Phys. Rev. D \textbf{69}, 105018 (2004)
[arXiv:hep-ph/0401045 [hep-ph]].

\bibitem{Pisarski:1993rf}
R.~D.~Pisarski,
Phys. Rev. D \textbf{47}, 5589-5600 (1993).



\end{thebibliography}
\end{document}